\documentclass[12pt]{article}
\usepackage{epsfig}
\usepackage{a4wide}

\input{epsf}

\newcommand{\be}{\begin{equation}} 
\newcommand{\ee}{\end{equation}}
\newcommand{\bea}{\begin{eqnarray}} 
\newcommand{\eea}{\end{eqnarray}}

\def\p{\partial}
\def\a{\alpha}

\def\l{\lambda}
\def\r{\rho}
\def\g{\gamma}
\def\th{\theta}
\def\z{\zeta}
\def\s{\sigma}
\def\ra{\rightarrow}

\def\R{{\cal R}}

\begin{document}

\title{Nonlinear spherical gravitational downfall of gas 
onto a solid ball: analytic and numerical results}

\author{Jos\'e Gaite\\
{\small\em Instituto de Matem{\'a}ticas y F{\'\i}sica
Fundamental,} 
{\small\em CSIC,}\\[-1mm] 
{\small\it Serrano 123, 28006 Madrid, Spain}\\and\\
Mari-Paz Zorzano\\
{\small\it Centro de
Astrobiolog{\'\i}a, CSIC-INTA,}\\[-1mm]
{\small\it Carretera de Torrej\'on a Ajalvir,}
{\small\it 28850 Torrej\'on de Ardoz, Madrid, Spain.}
}
\date{{\small May 14, 2003}}
\maketitle

{\abstract The process of downfall of initially homogeneous gas onto a
solid ball due to the ball's gravity (relevant in astrophysical
situations) is studied with a combination of analytic and numerical
methods.  The initial explicit solution soon becomes discontinuous
and gives rise to a shock wave.  Afterwards, there is a crossover
between two intermediate asymptotic similarity regimes, where the
shock wave propagates outwards according to two self-similar laws,
initially accelerating and eventually decelerating and vanishing, 
leading to a static state.  The numerical study allows one to
investigate in detail this dynamical problem and its time evolution,
verifying and complementing the analytic results on the initial
solution, intermediate self-similar laws and static long-term
solution.}

\vskip .2cm


\section{Introduction}

The problem of unsteady motion of compressible gas
giving rise to the formation of shock waves has been the subject of
continuing attention. In particular, the polytropic gas flow in one
dimension is amenable to powerful mathematical methods
\cite{Cour-F,LL,ChM}.  The introduction of gravitational forces 
complicates the problem but widens its range of applications.  

We study in this paper a dynamical problem that is very simple to
formulate: given a solid body with a spherically symmetric mass
distribution (a ball), placed in a homogeneous gas, analyze the
subsequent evolution of this gas under the body's gravity.  This
problem has applications in several astrophysical situations involving
gas accretion. For example, it could be a simplified model for the
formation of planetary atmospheres: as a result of gas accretion onto
the solid (rock) surface of a previously formed spherical core, an
atmosphere may form.  Alternatively, it could represent the downfall
of gas onto a neutron star.  The problem leads to solving the partial
differential equations of fluid dynamics in one dimension (the radial
distance). These equations are nonlinear and no general method of
solution is available. However, given the simplicity of the initial
and boundary conditions, many results can be obtained by purely
analytic means (as we shall see). The numerical integration of the
partial differential equations provides an alternative method of study
which is used here to validate and complement the analytic results.

Certain aspects of this fluid dynamical problem have been treated
before in the astrophysical literature \cite{Zel,KM}. However, a
complete treatment employing the full power of nonlinear fluid
dynamics methods has not yet been attempted. The thorough analytic and
numerical treatment provided here allows one to attain a picture of
the whole process and to connect with other problems in fluid
dynamics, especially, with gravitation.  In particular, the exact
description of the formation of the shock wave may illustrate general
features present in other processes.

This paper is organized as follows.  In section \ref{equations}, we
introduce the relevant magnitudes of the problem to get an intuitive
idea of the physics involved.  Afterwards, we introduce the fluid
equations to be used.  In section \ref{exact}, an exact solution is
found for the initial non-trivial dynamics, which occurs near the
ball's surface.  In the following sections we obtain two {\em
similarity} solutions valid for larger $t$, the first still confined
within a small height (section \ref{simil}), while the second is
valid for large radius (section \ref{simil2}). In section \ref{num}
we integrate numerically the partial differential equations of the
constant gravity and varying gravity case and compare the numerical
results with the analytical ones.  Finally, in section \ref{static},
we consider the long-time asymptotic static state and explore the
transition to it with the numerical solution.  In section
\ref{discuss} we summarize and discuss the results.

\section{Relevant magnitudes and fluid equations}
\label{equations}

The initial condition, namely, the (spherically symmetric) solid ball
in the homogeneous gas, is characterized by four parameters:
the radius $R$ and mass $M$ of the ball, and the pressure $P_0$ and
density $\r_0$ of the gas (we assume that the gas is perfect,
inviscid, and  polytropic).  In addition, we have the constant
of gravity $G$. From these five dimensional characteristic
parameters, we can form two independent dimensionless numbers:
the ratio of densities $(4/3)\pi R^3 (\r_0/M)$ and the ratio $R
P_0/(\r_0 G M)$. The quantity $P_0/\r_0 \sim c_0^2$ ($c_0$ being the
sound speed) is approximately the gas thermal energy per unit of
mass. On the other hand, $G M/R$ is the gas potential energy per unit
of mass on the ball. Their ratio measures the relative strength of
gravity in our problem.  To have any significant gas downfall, we must
demand that the ball's gravity dominates over the gas thermal energy,
that is, $R c_0^2/(G M) \ll 1$.  

Defining the scale at which the gas thermal 
energy is similar to its potential energy as $\R =  G M/c_0^2$, 
we demand that
\be
R \ll \R \ll \left({M\over \r_0}\right)^{1/3}.
\ee
The last length is the radius of a volume of gas such that
its mass is similar to $M$ and, therefore, we can
neglect the self-gravity of the gas.

Since the initial and boundary conditions
are spherically symmetric, 
we will assume that the whole process is spherically
symmetric and thus one-dimensional. 
We further consider an adiabatic evolution of the gas or, more
generally, a {\em polytropic} equation of state, 
$P \propto \rho^\g$ ($\g \geq 1$): 
for a perfect gas, with $\g = C_P/C_V$ (the constant ratio of 
specific heats), a polytropic process is adiabatic, but 
the notion of polytropic process is more general.
For numerical calculations, we will use $\g = 7/5$
(the adiabatic index of a perfect diatomic gas).
Then, we have the continuity equation,
the Euler equation and the thermodynamic equation:
\bea 
{\p\r\over\p t} + {\p(\r v)\over\p x} + {2(\r v)\over x + R}&=&0, 
\label{cont}\\ 
{\p v\over\p t} + v {\p v\over\p x} &=& -g(x) - {1 \over\r}
{\p P\over\p x},  \label{Eul}\\
{\p\over\p t}{P\over\r^\g} + v {\p \over\p x}{P\over\r^\g} &=& 0, 
\label{thermo}
\eea 
where $x$ is the radial distance from the ball surface and 
$g(x) = GM/(x+R)^2 = g R^2/(x+R)^2$ is the gravity acceleration
($g$ is the gravity at the ball surface). 
The initial conditions are: $\r(0,x) =
\r_0$, $P(0,x) = P_0$ and $v(0,x) = 0$ ($x \geq 0$). The boundary
condition at the solid surface is $v(t,0) = 0$.
We remark that the behaviour of the solutions of these equations 
for large $x$ is only determined by the initial conditions, 
which in some sense replace a boundary condition ``at infinity''.
However, in the numerical calculations we shall need a real boundary 
condition at some large value of $x$.

The thermodynamic equation (\ref{thermo})
(which expresses conservation of entropy or 
some more general thermodynamic quantity)
has the trivial solution $P/\r^\g = P_0/\r_0^\g$. 
Introducing the sound velocity $c$ ($c^2 = dP/d\r$), 
we can write the preceding solution as
\be
c^2(\r)/c_0^2 = (\r/\r_0)^{\g-1}\quad (c_0^2 = \g P_0/\r_0). 
\label{dens-c}
\ee
Therefore, we need to consider only the two remaining equations:
\bea 
{\p\r\over\p t} + {\p(\r v)\over\p x} + {2(\r v)\over x+R}&=&0, 
\label{cont2}\\ 
{\p v\over\p t} + v {\p v\over\p x} &=& -g(x) - {c^2(\r) \over\r}
{\p\r\over\p x}.  \label{Eul2}
\eea
However, in the presence 
of a shock discontinuity, the thermodynamic equation (\ref{thermo})
breaks down (as is most intuitive when this equation can be interpreted
as entropy conservation). Therefore, 
the solution (\ref{dens-c}) does not hold, 
as we shall see in the numerical solution.

The fluid equations can also be derived 
as conservation equations, namely, as equations for 
conservation of mass, momentum and energy. In our case, 
we need to take into account that $x$ is the radial distance and 
write the divergences in spherical coordinates; in particular, the 
divergence of the stress tensor $T^{ij} = \r
v^i v^j + g^{ij} P$ ($g^{ij}$ is the inverse metric).  
The fluid equations in conservative form become:
\bea
\frac{\partial \r}{\partial t}+
\frac{\partial [(x+R)^2 \r v]}{(x+R)^2 \partial x}&=&0,\\
\frac{\partial (\r v)}{\partial t}+
\frac{\partial \left[(x+R)^2 \r v^2\right]}{(x+R)^2 \partial x} +
\frac{\partial P}{\partial x}&=&-g(x) \r,\\
\frac{\partial (\r e + \r v^2/2)}{\partial t}+
\frac{\partial \left[(x+R)^2 (\r e + \r v^2/2 +P)v\right]}{(x+R)^2
\partial x} &=& -g(x) \r v.
\eea
They are equivalent to Eqs.\ (\ref{cont},\ref{Eul},\ref{thermo}), 
provided that the pressure is related with the internal energy 
(per unit mass) $e$ 
by $P=(\gamma-1)\rho e$ and that the solutions are smooth.
In the presence of a shock discontinuity, the conservation 
equations still hold, which is their main advantage. Their disadvantage
is that they are more complicated.

Defining the vector $\mathbf{U}=(\rho,\rho v,\rho e+\rho v^{2}/2)$,
we can express the equations in the following divergence form:
\be
\frac{\partial \mathbf{U}}{\partial t}+\frac{\partial
 \mathbf{F}(\mathbf{U})}{\partial x}=\mathbf{G}(\mathbf{U}),
\label{vdiverg}
\ee
with $\mathbf{F}(\mathbf{U})=(\rho v,\rho v^{2}+P,(\rho e+\rho
 v^{2}/2+P)v)$ and $\mathbf{G}(\mathbf{U})=(-2\rho v/(x+R),-2\rho v
^{2}/(x+R)-g(x)\rho,-2(\rho e +\rho v ^{2}/2+P)/(x+R) -g(x)\rho v)$
the source term. This form will be useful in the next section, and 
for numerical calculations as well.

\section{Early stage: exact solution}
\label{exact}

Initially, the gas will start falling with acceleration $g(x)$, that is,
with a negative velocity increasing as $v \sim -g(x)\,t$. However, it
will be stopped at the ball's surface, where the density and,
therefore, the pressure must increase.  Consequently, a wave
transmitting the boundary condition $v(t,0) = 0$ will propagate
outwards. It is therefore crucial to determine the law of propagation
of waves in the present conditions. This is called, in mathematical
terms, the analysis of {\em characteristics}. Once this is done, and
as long as the dynamics consists of the propagation of a {\em simple
wave}, one can obtain an exact solution. We do not have here the
opportunity to review the theory of simple waves, so we provide 
some indications and refer the
reader to general treatises on the theory of one-dimensional gas flow
for a comprehensive treatment \cite{Cour-F,LL,ChM}. 

For the moment, we confine ourselves to a spherical shell over the
surface of height much smaller than $R$, where we can consider
constant $g=GM/R^2$.  Hence, the problem becomes that of the fall of
gas on a flat surface (the ground).  So we must use the
one-dimensional form of the vector divergence and, therefore, we must
neglect the last term on the left-hand side of the continuity equation
(\ref{cont2}), writing it as
\be 
{\p\r\over\p t} + {\p(\r v)\over\p x} =0. \label{cont1} 
\ee

The gas still unperturbed
by the wave merely falls with velocity $v=-g\,t$. In a reference
system falling with it, it is at rest and, therefore, the speed of sound
is $c_0$. Hence, in the original reference system the wave front is
located at $x_f = c_0 t - g\, t^2/2$. Then, for $x > x_f$, the
solution is trivial, and we only have to find the solution for $x < x_f$.  
In the falling frame, the acceleration of gravity vanishes and the 
set of two equations (\ref{cont1}) and (\ref{Eul2}) reduces to a 
``gas tube problem" that can be solved by introduction of the Riemann
invariants \cite{Cour-F,LL,ChM}. 

The Riemann invariants are best introduced by writing 
the equations in the divergence form (\ref{vdiverg}). 
The replacement of spherical geometry with cartesian geometry 
reduces the source term to 
$\mathbf{G}(\mathbf{U})=(0,-g\rho,-g\rho v)$ and, furthermore,
the change to the falling frame removes it.
Then we can write the divergence form (\ref{vdiverg}) 
as a quasilinear partial differential equation:
\be
\frac{\partial \mathbf{U}}{\partial t} + A(\mathbf{U})\cdot
\frac{\partial \mathbf{U}}{\partial x}=0,
\ee
where $A(\mathbf{U})$ is the Jacobian matrix of derivatives of 
$\mathbf{F}(\mathbf{U})$ with respect to $\mathbf{U}$. 
The solution of this equation is carried out by diagonalizing $A$:
the eigenvalues $\l$ of $A$ 
(which are real because the equations are hyperbolic)
define the {\em characteristic curves}
\be
\frac{dx}{dt}= \l(\mathbf{U}) ,
\ee
and the left eigenvectors of $A$ (the eigenvectors of its 
transpose) define the Riemann invariants $J$, that is, functions
that satisfy
\be
A(\mathbf{U})^T\cdot
\frac{\partial J}{\partial \mathbf{U}}=
\l \frac{\partial J}{\partial \mathbf{U}}
\ee
for some eigenvalue $\l$. It is easy to prove that $J$ is constant 
along the characteristic associated with the corresponding $\l$. 
The characteristics are the directions of propagation of disturbances 
(waves). Since we have three eigenvalues, namely, $v$ and $v \pm c$,
we have that the first family of characteristics are the streamlines and 
the other two families are forward or backward characteristics, 
according to the sign. 

\begin{figure}
\centering
  \epsfxsize=10cm 
\epsfbox{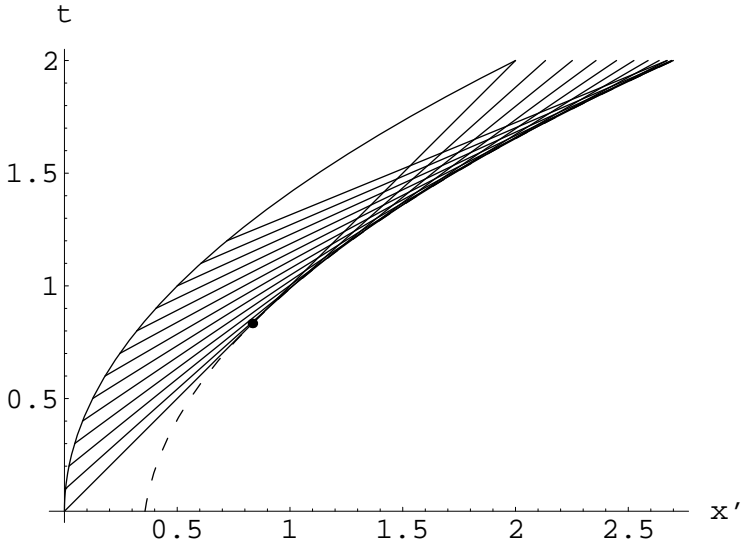}
\caption{\label{Fig1} Line described by the ground and 
forward characteristics
in the falling frame (in units such that $c_0 = g = 1$ and for
$\g=7/5$).  The dashed line $(\p x'/\p v')_t=0$ is the envelope of the
characteristic lines.  At the cusp point $(5/6,5/6)$, the
characteristics cross and the solution becomes multivalued.}
\vskip .2cm 
\end{figure}

Let us denote $x',v',c'$ the coordinate and variables in the falling
frame.  Initially we have a {\em constant state} (with $v'$ and $P$
constant), which is preserved for $x' \geq c_0 t$. Therefore, the backward
characteristics crossing the forward characteristic $x' = c_0 t$
transmit a constant value of the Riemann invariant $J_-$, so the
solution is a forward simple wave. Given that
$$J_- = v'  - \frac{2c'}{\g-1} = - \frac{2c_0}{\g-1}\,,$$
the sound velocity is related 
with the gas velocity by $c' = c_0 + (\g-1)v'/2$. 
Furthermore, the fact that the solution is a forward simple wave
implies that the forward characteristics are straight lines 
(see Fig.\  \ref{Fig1}).
Hence, the wave's propagation law $v'=F[x'-(v'+c')\,t]$ is an
implicit equation for $v'$, 
assuming that we can determine the function
$F$. This is done using the boundary condition $v(t,0) = 0$. The
implicit equation is algebraic and its solution is
straightforward. We then obtain: 
\be 
v(x,t) =\left\{ {\begin{array}{l}
{\displaystyle -gt, \quad x \geq c_0 t - {g\over 2} t^2}\\
{\displaystyle  -{c_0\over \g} - {\g-1\over 2\g} gt + \sqrt{\left({c_0\over
      \g} + {\g-1\over 2\g} gt\right)^2 - {2g\over\g}x}, \quad
           x \leq c_0 t - {g\over 2} t^2.} \end{array}} \right.
\label{vel}
\ee
For a simple wave, the density is a definite function of the velocity
\cite{LL}. We can calculate it from Eq.\ (\ref{dens-c}) to be
\be
\r(x,t) = \r_0 \left[1+{\g-1\over 2}\,{v(x,t)+gt\over
c_0}\right]^{{2\over\g-1}}.     \label{dens}
\ee
These expressions for $\r$ and $v$ constitute 
the solution of Eqs.~(\ref{Eul2}) 
and (\ref{cont1}) with the given boundary conditions. 

We note a peculiarity of the solution found above: the wave
front's coordinate $x_f = c_0 t - g t^2/2$ begins increasing but, for
$t > c_0/g$, turns to decreasing and, eventually, returns to the
origin (a free fall with initial velocity $c_0$).  On the other hand,
a shock wave singularity occurs before the turning point:
One can calculate $(\p x/\p v)_t$, and its null locus defines a line
in the $xt$-plane, namely, $x = [2c_0 + (\g-1)g t]^2/(8\g
g)$. Initially, the corresponding $x$ is larger than $x_f$, so it is
unphysical, but, for $t=2c_0/[(\g+1)g]$, it meets the wave front; that
is, the falling gas overtakes the gas just below it and a shock wave
arises. The relevant geometry is represented in Fig.~1, in the
falling reference frame. From the moment of formation of the shock 
onwards, the solution
(\ref{vel},\ref{dens}) ceases to be valid. Furthermore, the analysis
of the propagation of a shock wave
cannot be done in terms of simple waves.

\section{Similarity solution with constant gravity}
\label{simil}

When the initial conditions are sufficiently simple, the solution of a
one-dimensional gas flow problem may adopt a self-similar structure
and be greatly simplified by writing the equations in the appropriate
variables \cite{LL,Sedov}, in which the partial differential equations 
become ordinary differential equations (ODE). 

In our case, a self-similar solution is not possible, 
since we have too many parameters in the initial conditions.
Nevertheless, it is commonly observed that nonlinear 
equations that do not have similarity solutions at the outset 
develop them in an {\em intermediate asymptotic} regime, that is, 
a regime between two very different scales where both scales can be
neglected \cite{Bar}. 
We shall see that we can find a similarity solution of this kind.

In the equations with constant gravity, namely,
(\ref{Eul}), (\ref{thermo}) and (\ref{cont1}),
$R$ and $GM$ only appear in the combination $g=GM/R^2$;
furthermore, the only parameter in the initial and boundary conditions is 
$\r_0$. Hence, we have one less parameter. 
However, we can still form the two {\em independent} dimensionless
variables $gt/c_0$ and $gx/c_0^2$, so we will have to remove another
parameter to have a similarity solution.  

If we assume that the initial gas temperature is very low, then $c_0
\rightarrow 0$ and the only independent dimensionless variable is 
$\xi = x/(g t^2)$  (we measure the distance 
$x$ from the ground). This corresponds to the intermediate asymptotics
$c_0^2/g \ll x \ll R$; equivalently, in terms of the dimensionless
height variable $x^* = gx/c_0^2$, it corresponds to 
$1 \ll x^* \ll  R^* = \R/R$ (recall that $\R \gg R$). 

To take into account the presence of shock waves and the consequent
dissipation, we consider the full set of equations 
(\ref{Eul}), (\ref{thermo}) and (\ref{cont1}). 
We can express them in 
dimensionless form by introducing new variables:
\be
v = {x\over t} u, \quad \r = \r_0 r, \quad P = {x^2\over t^2}\,\r_0\,p.
\ee
Some straightforward algebra then yields,
\bea
\xi[u' + (u-2) {r'\over r}] + u &=& 0,\\
(u-2)\,\xi u' + u^2 - u &=& -\xi^{-1} -r^{-1}(2p + \xi p') ,\\
(u-2)\,\xi [\ln(p/r^\g)]' + 2(u-1) &=& 0.
\eea
Upon making the change of variables 
\be
\left\{
      {\begin{array}{l}
  {\displaystyle \xi = \tilde\xi -{1\over 2},}\\
  {\displaystyle u = {\tilde\xi\tilde u -1\over \xi},} \\
  {\displaystyle  p =  {\tilde\xi^2 \tilde p\over \xi^2}} \end{array}} 
\right.
\label{change}
\ee
(corresponding to changing to the falling reference frame),
the  $\xi^{-1}$ term in the second equation vanishes and 
the system of ordinary differential equations 
simplifies to 
\bea
\tilde{\xi}[\tilde{u}' + (\tilde{u}-2) {r'\over r}] + \tilde{u} &=& 0,\\
(\tilde{u}-2)\,\tilde{\xi} \tilde{u}' + \tilde{u}^2 - \tilde{u} &=&
-r^{-1}(2\tilde{p} + \tilde{\xi} \tilde{p}') ,\\
(\tilde{u}-2)\,\tilde{\xi} [\ln(\tilde{p}/r^\g)]' + 2(\tilde{u}-1) &=& 0.
%\label{sys}
\eea
Similarity equations of this type have also been studied in Ref.\
\cite{Zel}. However, the authors do not make the
change to the falling reference frame, and restrict themselves to
solving the equations by a series expansion near the ground. Instead,
we follow here the general methods exposed in Ref.\ \cite{Sedov}.

In terms of the variable $\tau = \ln \tilde{\xi}$, the previous
equations imply a system of autonomous nonlinear ODE, namely,
\bea
{d\tilde{u}\over d\tau} &=& {\frac{\tilde{p}(\tau)\,\left[ 2 +
      \g\,\tilde{u}(\tau) \right] - r(\tau)\,\tilde{u}(\tau)\left( 2 -
      3\,\tilde{u}(\tau) + {{\tilde{u}(\tau)}^2} \right) }
      {-\g\,\tilde{p}(\tau) + r(\tau)\,{{\left[ \tilde{u}(\tau) - 2
      \right] }^2}}}, \label{ODE1}\\
{dr\over d\tau} &=&  r(\tau){\frac{-2\,\tilde{p}(\tau) + 
       r(\tau)\,\left[ \tilde{u}(\tau) -2 \right] \,\tilde{u}(\tau)}{
     \left( -\g\,\tilde{p}(\tau)   + 
       r(\tau)\,{{\left[  \tilde{u}(\tau) -2 \right] }^2} \right) \,
     \left[\tilde{u}(\tau)-2 \right] }}, \label{ODE2}\\
{d\tilde{p}\over d\tau} &=&  \tilde{p}(\tau){\frac{2\,\g\,\tilde{p}(\tau) - 
        r(\tau)\,\left( 4 - \left( 6 + \g \right) \,\tilde{u}(\tau) + 
           2\,{{\tilde{u}(\tau)}^2} \right)}{-\g\,\tilde{p}(\tau) + 
      r(\tau)\,{{\left[ \tilde{u}(\tau)-2 \right] }^2}}}.
\eea
Given the homogeneity properties of these equations with respect to
$\tilde{p}$ and $r$, it proves convenient to introduce the variable $\th=
\tilde{p}/r$ (the dimensionless form of the temperature for the perfect
gas with pressure $P$ and density $\r$). Thus,
\be
{d\th\over d \tilde{u}} = \th\,{\frac{2\,\th\,\left( 1 +
     \g\,(\tilde{u}-2) \right) - \left( \tilde{u} - 2\right) \, \left(
     4 - \left( 5 + \g \right) \,\tilde{u} + 2\,{{\tilde{u}}^2}
     \right) }{\left( \tilde{u} - 2 \right) \, \left( \th\,\left( 2 +
     \g\,\tilde{u} \right) - \tilde{u}\,\left( 2 - 3\,\tilde{u} +
     {{\tilde{u}}^2} \right) \right) }} 
\label{ODE}
\ee
The solution of this equation provides the relation between the
``velocity'' $\tilde{u}$ and the ``temperature'' $\th$,
$\th(\tilde{u})$.  Substituting it back into Eq.~(\ref{ODE1}), we have
an ODE for $\tilde{u}(\tau)$, which is immediately solved by a
quadrature. Analogously, one solves for $r(\tau)$.

We must now translate the initial and boundary conditions of the
original partial differential equations (PDE) into an initial
condition for the ODE (\ref{ODE}): at $t=0$, $\r(0,x) = \r_0$, $P(0,x)
= P_0 = 0$ ($c_0 \ra 0$), and $v(0,x) = 0$, implying that $r=1$ and
$u=0$ for $\tilde\xi = \xi = \infty$; at $x=0$, $v=0$, which implies,
since $v = gt(\tilde\xi \tilde u-1)$, that $\tilde u =2$ for $\xi=0$
and $\tilde\xi = 1/2$. The problem seems overdetermined. However, 
recalling the exact solution of section \ref{exact}, which gives
rise to a discontinuity (shock wave), we expect that 
the self-similar solution is discontinuous, so that there is no 
contradiction (in fact, this happens in most self-similar solutions
\cite{LL,Sedov}).

When discontinuities arise, the original differential equations 
break down, but they can be expressed as integrals over test functions,
which may have discontinuous solutions ({\em weak solutions}) 
\cite{Cour-F,LL,ChM}. Integration of Eq.\ (\ref{vdiverg}) across a 
discontinuity yields the Rankine-Hugoniot jump conditions 
 $\mathbf{F}(\mathbf{U_1})= \mathbf{F}(\mathbf{U_2})$.
These conditions can be written as
\be
\left\{
      {\begin{array}{l}
         \r_1 v_1 = \r_2 v_2\,,\\
         \r_1 v_1^2 + P_1 = \r_2 v_2^2 + P_2\,, \\
   {\displaystyle {v_1^2\over 2}  + {P_1 \over \r_1} +  e_1
         = {v_2^2\over 2} + {P_2\over \r_2} +  e_2\,,} \end{array}} 
\right.
\label{RH}
\ee
where subscripts refer to the values on each side of the shock and
$e$ is the internal energy;  
for a perfect gas, $e = P/[\r (\g-1)]$.  These equations hold in a
coordinate system in which the shock surface is at rest. On the other
hand, the location of the shock must be at fixed $\xi$, say $\xi_s$
(the subscript $s$ will denote variables pertaining to the shock motion).
Consequently, the shock wave velocity is $$v_s = {dx_s \over dt} =
\xi_s g {dt^2 \over dt} = 2\xi_s g t = 2 {x_s \over t},$$ that is,
$u_s = 2$ (and also $\tilde u_s = 2$). Then,
\be
\left\{
      {\begin{array}{l}
  {\displaystyle r_1 (u_1 -2) = r_2 (u_2 -2),}\\
  {\displaystyle p_1 + r_1 (u_1-2)^2 = p_2 + r_2 (u_2-2)^2,} \\
  {\displaystyle {(u_1-2)^2\over 2} + {\g \over \g-1}{p_1 \over r_1} = 
       {(u_2-2)^2\over 2} + {\g \over \g-1}{p_2 \over r_2},}\end{array}} 
\right.
\label{shock2}
\ee
valid in both the rest and the falling reference frames.
Using the falling frame and eliminating $r_2/r_1$ between the first
and second equations,
\be
\left\{ {\begin{array}{l} 
{\displaystyle {\th_1\over \tilde{u}_1-2} + \tilde{u}_1-2 =
      {\th_2\over \tilde{u}_2-2} + \tilde{u}_2-2,} \\
{\displaystyle (\tilde{u}_1-2)^2 + {2\g \over \g-1}\th_1 = 
       (\tilde{u}_2-2)^2 + {2\g \over \g-1}\th_2.} \end{array}} \right.
\label{shock3}
\ee

We have obtained an involutive mapping of the plane $(\tilde u,\th)$
as the relation between the variables at either side of the shock
discontinuity. Since we have the condition that as $x \rightarrow
\infty$ we go to $(0,0)$ and (in the falling frame) 
these values hold all the way down to the
shock surface, we just need the image of the origin under the
mapping. This yields the point
$\left(4/(\g+1), 8\left(\g-1 \right)/{\left(\g+1 \right) }^2\right)$. 
Therefore, we need the solution of Eq.~(\ref{ODE}) that goes through
this point. We can see in the example of Fig.~2 ($\g = 7/5$) 
that the strand of this
solution that departs from (5/3,5/9) 
ends at the point (2,0), hence satisfying the
boundary condition at $x=0$.% 
\footnote{Notice that this implies that the asymptotics
$c_0 \rightarrow 0$ is of the {\em first kind} (the simple case) 
according to the denomination of Barenblatt \cite{Bar}.}

Now, substituting the function just computed back into Eq.~(\ref{ODE1}), we
obtain
\be
d\tau = {\frac{\g\,\th(\tilde{u}) - {{\left( \tilde{u} - 2 
     \right) }^2}}{-\th(\tilde{u})\,\left( 2 + \g\,\tilde{u}
     \right) + \tilde{u}\,\left( 2 - 3\,\tilde{u} +
     {{\tilde{u}}^2} \right) } }\,d\tilde{u},
\ee
from which we can deduce the interval of $\tau$ between the points 
$\left(4/(\g+1), 8\left(\g-1 \right)/{\left(\g+1 \right) }^2\right)$ 
and (2,0) by integration. The result for $\g = 7/5$
is 0.0880104. Hence, 
we obtain the quotient between 
the corresponding values of $\xi$. This quotient gives 
the location of the shock relative to the ground:
${\tilde\xi}_s/(1/2) = \exp 0.0880104 = 1.09200 \Rightarrow 
\xi_s = 0.0919995/2 = 0.0459997$, so that the coordinate of the shock is 
$x_s = 0.0459997 g t^2$. 

\begin{figure}
\centering
  \epsfxsize=8cm 
\epsfbox{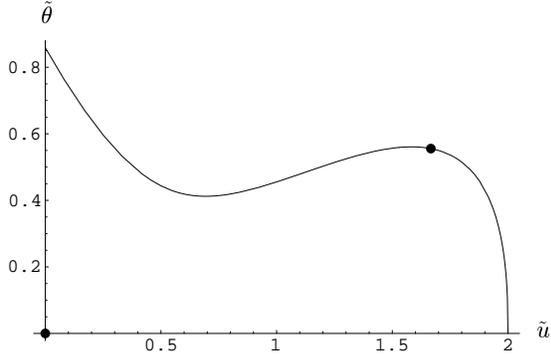}
\vskip -.4cm 
\caption{\label{Fig2} Relevant solution of the ODE (\ref{ODE}) for $\g
= 7/5$, displaying the origin and its image at the shock surface, the
point $(5/3,5/9)$.}
\vskip .2cm 
\end{figure}

Once we have described the self-similar solution, let us analyze in
detail its range of validity.  The self-similar solution is valid for
$c_0 \rightarrow 0$, that is, for the dimensionless variables $t^{*} =
gt/c_0$ and $x^{*} = gx/c_0^2$ sufficiently large.  Moreover, $x^{*}
\ll R^*$, but the upper limit of the $t^{*}$ asymptotics is still
indefinite.  The remaining condition can be found by using the
dimensionless variables $\xi = x^{*}/(t^{*})^2 = x/(g t^2)$ and $\z =
x^{*}/t^{*} = x/(c_0 t)$, such that the former only depends on $g$ and
the latter on $c_0$.  We have that $c_0 \rightarrow 0$ implies that
$\z \gg 1$.  To realize the consequences of the previous conditions it
is convenient to consider the logarithmic variables $\ln x^{*}$ and
$\ln t^{*}$, on the one hand, and the linearly related variables $\ln
\xi$ and $\ln \z$, on the other hand.  The three equations $\ln x^{*}
= \ln R^*$, $\ln t^{*} = 0$ and $\ln \z = 0$ define a triangle in
the $(\ln x^{*},\ln t^{*})$ plane in which the complete intermediate
asymptotics holds; to be precise, in the interior region far from the
borders.  Moreover, part of this region corresponds to the trivial
solution above the shock (the part with $\xi > \xi_s$).

\section{Similarity solution with variable gravity}
\label{simil2}

For heights of the order of the ball radius or larger, we
have to return to the full continuity equation (\ref{cont})
and reset $g(x) = GM/(x+R)^2$ in the Euler equation (\ref{Eul}). 
We have an additional parameter, 
namely, the radius $R$. If we further assume that $R \ra 0$, 
we can find a similarity solution, that is, in
the intermediate asymptotics $R \ll x \ll \R$ ($R^* \ll x^* \ll (R^*)^2$). 
Now, the dimensionless variable $\xi$ becomes $\xi = x^3/(GM t^2)$.  
Hence, we can express the continuity and Euler equations as
\bea
\xi[3u' + (3u-2) {r'\over r}] + 3u &=& 0,\\
(3u-2)\,\xi u' + u^2 - u &=& -\xi^{-1} -r^{-1}(2p + 3\xi p'). \label{DE}
\eea
The energy equation becomes
\be
(3u-2)\,\xi [\ln(p/r^\g)]' + 2(u-1) = 0.
\ee

These three equations look similar to the ones corresponding to 
constant gravity, but they are more difficult to analyze:%
\footnote{An extensive study of self-similar spherical accretion
with an initial power-law density distribution is provided by 
Ref.\ \cite{KM}.}
the change of variables that transformed the latter into 
an autonomous system is not available.
In fact, the free-fall 
(pressureless) problem is now nontrivial, but it can be solved 
(see Appendix). The solution, in parametric form, reads
\bea
\xi &=& {8\cos^{6}\alpha \over  2\alpha +\sin(2\alpha)}, \\
u &=& -(\alpha + {\sin(2\alpha)\over 2})\,\frac{\sin\alpha}{\cos^{3}\alpha},\\
r &=& {\r\over \r_0} = \frac{8\,\cos^{-3}\a}
  {9\,\cos\a - \cos (3\,\a) + 12\,\a\,\sin \a}, 
\eea
where $\alpha \in (0,\pi/2)$. 
The corresponding graphs for $u$ and $r$ are shown in Fig.~3.

\begin{figure}
\centering
  \epsfxsize=8cm 
\epsfbox{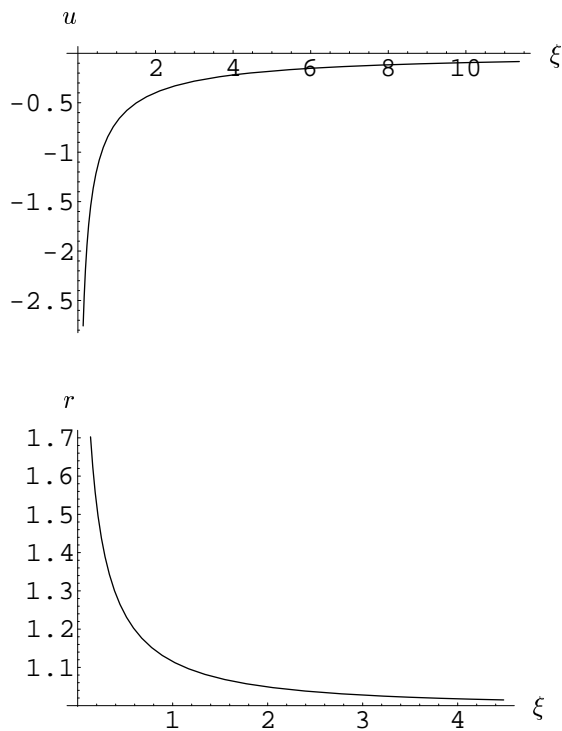}
\flushleft
\vskip -.5cm 
\caption{\label{Fig3} Similarity solution $\{u(\xi),r(\xi)\}$ 
for free fall with varying gravity.}
\vskip .2cm 
\end{figure}

As regards the complete similarity solution, it must match the free
fall solution outside a spherical shock wave with the inner solution
with pressure.  In turn, this solution should be matched with the
previous solution near the ground. 
These matchings are difficult to obtain analytically, so we will
rather rely on numerical solutions (next section).
At any rate, we deduce that, 
as the shock moves from $x_s\ll R$ to $x_s \gg R$, its velocity
experiences a crossover from $v_s = 2 x_s/t = 2 \xi_s g t$ to $v_s =
2/3\,(x_s/t) = 2/3\,(GM\xi_s)^{1/3} \,t^{-1/3}$ (with {\em different}
$\xi_s$), changing from accelerating to decelerating.   
In addition, the shock must disappear 
when $t \gg GM/c_0^3$  and $x_s \gg \R$ (see section \ref{static}).

Of course, a rough analysis of the range of validity of this
similarity solution is also possible.  It is valid for $R \rightarrow
0$, that is, for the dimensionless variables $\hat{x} = x/R$ and
$\hat{t} = \sqrt{GM}/R^{3/2}\,t$ sufficiently large.  It is convenient
to use the dimensionless variables $\xi = \hat{x}^3/\hat{t}^2$ and $\z
= \hat{x}/\hat{t} = \sqrt{R/GM}\,x/t$, such that the former does not
depend on $R$.  We have that $R \rightarrow 0$ implies that $\z \ll
1$.  Considering the logarithmic variables $\ln \hat{x}$ and $\ln
\hat{t}$, on the one hand, and the linearly related variables $\ln
\xi$ and $\ln \z$, on the other hand, the three equations $\ln \hat{x}
= 0$, $\ln \hat{t} = 3/2\, \ln (\R/R)$ and $\ln \z = 0$ define the
triangle wherein the complete intermediate asymptotics holds (far from
the borders).

\section{Numerical solution}
\label{num}

In this section we will numerically validate and complement
the main results obtained above. As was mentioned in 
section \ref{equations}, 
our aim is to integrate the equation 
\be
\frac{\partial \mathbf{U}}{\partial t}+\frac{\partial
 \mathbf{F}(\mathbf{U})}{\partial x}=\mathbf{G}(\mathbf{U})
\label{divform}
\ee
for the cases of constant and varying gravity.
 Relying on the
 operator splitting approach, we integrate first the conservative part
 of the equation using the Lax numerical scheme and then integrate
 the source term:
\be
\mathbf{U}_{j}^{n+1}=\frac{\mathbf{U}_{j+1}^{n} +
 \mathbf{U}_{j-1}^{n}}{2} - \Delta t\,
\frac{\mathbf{F}(\mathbf{U}_{j+1}^{n}) -
 \mathbf{F}(\mathbf{U}_{j-1}^{n})}{2\,\Delta x} + \Delta t\
 \mathbf{G}(\mathbf{U}_{j}^{n}).
\ee
With this we get a discrete approximation of the value of $\mathbf{U}$
at $x_j=j\Delta x$ and time $t=n\Delta t$ to second order accuracy,
$\mathbf{U}_j^{n}$.  We will show the results in dimensionless
variables $x^{*}=x/(\frac{c_0^{2}}{g})$ and $t^{*}=t/(\frac{c_0}{g})$ and
 fix $\Delta x^{*}=0.005$ and $\Delta t^{*}=\frac{\Delta
x^{*}}{20}$.  We have chosen $\Delta t$ so that the
Courant-Friedrichs-Lewy stability condition of the scheme is satisfied
\cite{Num}.
This condition esentially states that we are required to choose a time
step smaller than the smallest characteristic physical time in the
problem, $\Delta t^{*}\leq \frac{\Delta x^{*}}{|v_{max}^{*}|}$, which
is the time for the velocity to lead to a flow over a distance $\Delta
x^{*}$ (during our integration time $v^{*}=v/c_0 \leq
v_{max}^{*}=20$).  We integrate the problem with initial condition
$\rho(x,0)=1$, $P(x,0)=\gamma c_0^{2}\rho_0$, $v(x,0)=0$, adiabatic
index $\gamma=7/5=1.4$ and boundary condition at the solid surface
$v(t,0)=0$.

\begin{figure}[h]
\begin{center}
\epsfig{figure=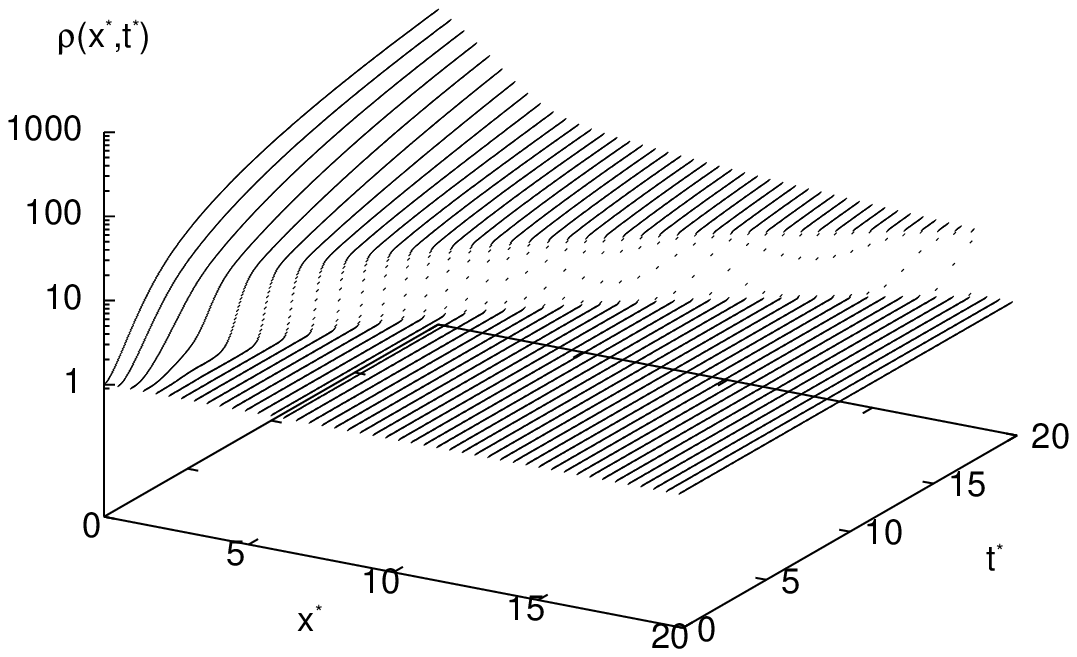,height=8.5cm,width=7.5cm}
\epsfig{figure=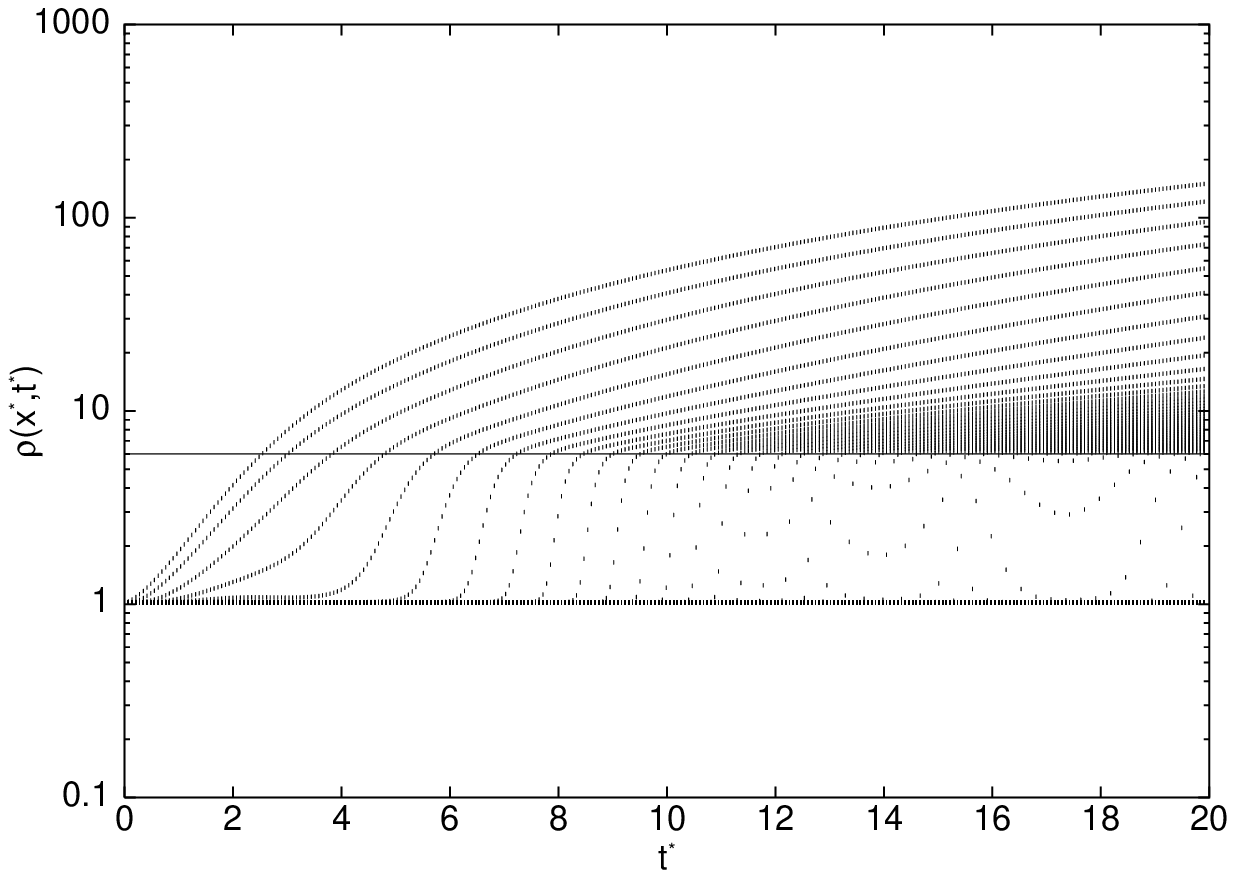,height=7.5cm,width=7.5cm}
\caption{\label{gcte} Density distribution and time evolution (for
constant $g$) in (decimal) logarithmic scale. Three-dimensional view
(left) and lateral view (right). The density ratio across the shock
tends to $\rho_2/\rho_1=(\gamma+1)/(\gamma-1)=6$ as expected for a
polytropic gas.}
\end{center}
\end{figure}

In the case of constant gravity, we can rewrite the equations of gas
 downfall on a flat surface (\ref{Eul2}) and (\ref{cont1}) in the
 divergence form (\ref{divform}) using
 $\mathbf{G}(\mathbf{U})=(0,-g\rho,-g\rho v)$ (as was mentioned in
 section \ref{exact}).  We show in Fig.\ \ref{gcte} (left) the density
 evolution in (decimal) logarithmic scale for the case of constant
 gravity. For the initial stages up to
 $t^{*}=\frac{2}{\gamma+1}\approx 0.83$, before the shock wave
 develops, the density and velocity values are coincident with those
 given by the analytical solutions Eqs.  (\ref{vel}) and (\ref{dens})
 (section \ref{exact}). Then the shock formation induces a
 discontinuity in the density, pressure and velocity distribution and
 the analytic solution is no longer valid. For a polytropic gas and
 based on the Rankine-Hugoniot conservation conditions accross the
 shock, one can see that the ratios $v_2/v_1$ and $P_2/P_1$ can be
 arbitrarily large, whereas (for these strong shocks) when
 $P_2/P_1\rightarrow \infty$, the density ratio $\rho_2/\rho_1$ tends
 to the constant limit $\rho_2/\rho_1=(\gamma+1)/(\gamma-1)$
 \cite{LL}. In our case with $g$ constant, the pressure and velocity
 variation across the shock continue to grow as long as the shock
 exists, while the discontinuity in the density tends to the limiting
 value $\rho_2/\rho_1=6$ ($\gamma=1.4$), as is shown in the lateral
 view of the density in Fig.\ \ref{gcte} (right).

\begin{figure}[h]
\begin{center}
\epsfig{figure=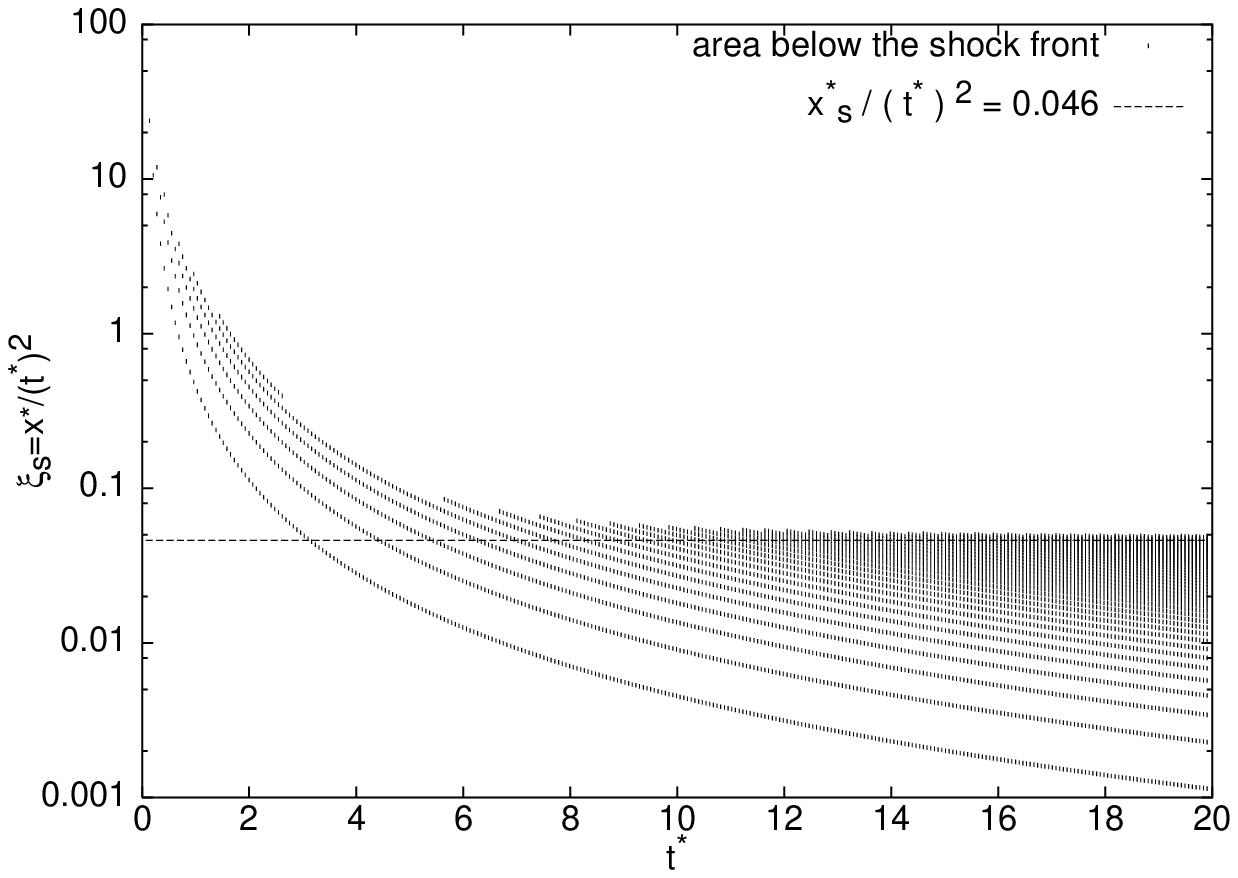,height=7.5cm,width=7.5cm}
\epsfig{figure=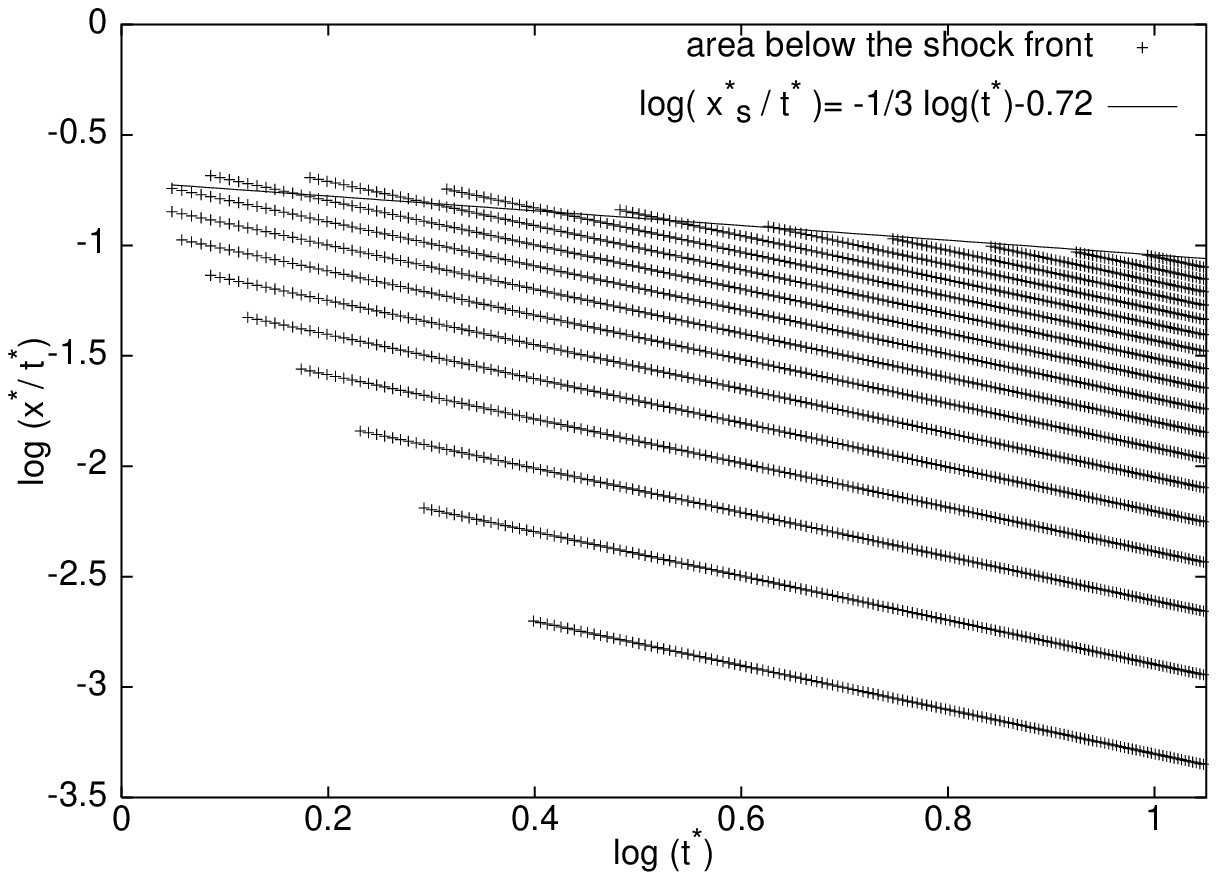,height=7.5cm,width=7.5cm}
\caption{\label{xi} Self-similar evolution. We plot, for all the
points $x^{*}$ between the surface and the shock front, the time
evolution of: (left) $\xi=x^{*}/ (t^{*})^{2}$ for the case of constant
gravity and (right) $(x^{*}/t^{*})$ in logarithmic scale for the case
of varying gravity with $R^{*}=10$. (Left) If gravity does not change
with height, the value of $\xi$ at the shock front (upper line) tends
to a constant value $\xi_s=0.046$. (Right) When gravity changes with
height, at the shock front (upper line) we can fit
$x_s^{*}/t^{*}=(t^{*})^{-1/3}\exp{(-0.72)}$ and therefore
$\xi_s=\frac{(x_s^{*})^{3}}{(t^{*})^{2}(R^{*})^{2}}=0.0011$. Both
results are in agreement with the analytic predictions of self-similar
solutions.}
\end{center}
\end{figure}

Following the asymptotic similarity analysis we expect the shock
  position $x_s$ to move as $x_s\approx \xi_s g t^{2}$, for the case
  of constant gravity (section \ref{simil}), or equivalently
  $x_s^{*}=\xi_s(t^{*})^{2}$.  In Fig.\ \ref{xi} (left) we plot all
  the points $x^{*}$, in the area affected by the wave
  ($x^{*}|\rho(x^{*},t^{*})>\rho_0$) rescaled by $(t^{*})^{2}$. The
  shock front is at the upper limit of this area. The value
  $\xi=x_s^{*}/ (t^{*})^{2}$ (evaluated at the shock) evolves to a
  constant value $\xi_s=0.046$ and therefore, beyond a transient time,
  the shock moves with constant acceleration. This is in agreement
  with the analytic prediction.

Next we study the time evolution of this process with spherical
  symmetry for the case of varying gravity, with
  $\mathbf{G}(\mathbf{U})=(-2\rho v/(x+R),-2\rho v
  ^{2}/(x+R)-g(x)\rho,-2(\rho e +\rho v ^{2}/2+P)/(x+R) -g(x)\rho v)$,
  $g(x)=g R^{2}/(x+R)^{2}$ and $g$ the gravity at the surface of the
  ball.  For our study case we choose
  $R^{*}=R/(\frac{c_0^{2}}{g})=10$.   We restrict ourselves to
  this relatively small value because the magnitude of the
  discontinuity at the shock and the velocity of the downfalling gas
  increase with $R^{*}$. So the spatial and temporal grid sizes must be
  fixed sufficiently small to be able to handle these sharp
  discontinuities and satisfy the Courant-Lewy condition for maximal
  velocities in the grid. This in turn imposes a limitation on the
  maximal $R^{*}$ to be considered.

We remark that, of course, the numerical integration
has to be restricted to a finite interval of $x^*$, namely
$[0,x^*_{max}]$, and therefore we need a boundary condition at
$x^*_{max}$.  This is not a problem in the constant $g$ case, because
what happens for large $x^*$ is {\em exactly} known: the gas is in
free fall with acceleration $g$ and $\r = \r_0$ (this is a valid
boundary condition as long as the shock front has not reached
$x^*_{max}$).  In contrast, in the varying $g(x)$ case, we impose as
boundary condition the initial values $\rho(t,x^{*}_{max})=\rho_0$ and
$P(t,x^{*}_{max})=P_0=\gamma c^{2}_0\rho_0$, which are similar
to the real values if $x^{*}_{max} \gg (R^*)^2$ and as long as
the shock front is sufficiently far from $x^*_{max}$.  For our
calculations we use $x^{*}_{max} = 450$.

  In Fig.\ \ref{xi} (right) we plot $x^{*}/t^{*}$
  vs. $t^{*}$ in (natural) logarithmic scale. The lower area is the
  area between the shock front and the surface. As one can see in the Figure, 
we can fit the motion of the shock front
  as $x_s^{*}/t^{*}=(t^{*})^{-1/3}\exp{(-0.72)}$. We get the
  corresponding value of
  $\xi_s=\frac{(x_s^{*})^{3}}{(t^{*})^{2}(R^{*})^{2}}=0.0011$.  Over
  this short range of distances and time, we are already able to find
  the asymptotic behaviour predicted in section \ref{simil2} in spite
  of the strong conditions needed for the derivation of the similarity
  solution $c_0 \rightarrow 0$ and $R \rightarrow 0$ which are hardly
  satisfied in this particular case (for these cases $v$ is not so
  large as compared to $c_0$, see Fig.\ \ref{evol} for a more detailed
  view of the dynamics of the main physical quantities).


\section{Transition to the static state}
\label{static}

The dissipation associated with the shock wave 
makes us expect that the gas
must reach a stationary state with more
entropy than the initial state. 
The properties of the stationary state are easily studied through 
the hydrostatic equation, 
$$\int {dP\over \r} = -\int g \,dx,$$ with $P \propto \r^\g$ and with
either constant or variable gravity $g (x)= GM/(x+R)^2$.  It is easy
to obtain the solution in the latter case (the constant $g$ solution
can be obtained expanding this solution in $x/R$ or directly):
\be
\left({\r\over \r_b}\right)^{\g-1} = 1 - {\g-1\over \g}{\r_b g_b R\over P_b} 
\left(1-{R\over x+R}\right),
\ee
where the subscript $b$ denotes values at the ball's surface.
It is more convenient to take the reference at
infinity, where the density and pressure keep their initial values, so that
\be
\left({\r\over \r_0}\right)^{\g-1} = 
1 + {\g-1\over \g}{\r_0 \over P_0}\,{G M\over x+R}  = 1 + (\g-1){\R\over x+R}.
\label{law}
\ee

\begin{figure}[h]
\begin{center}
\epsfig{figure=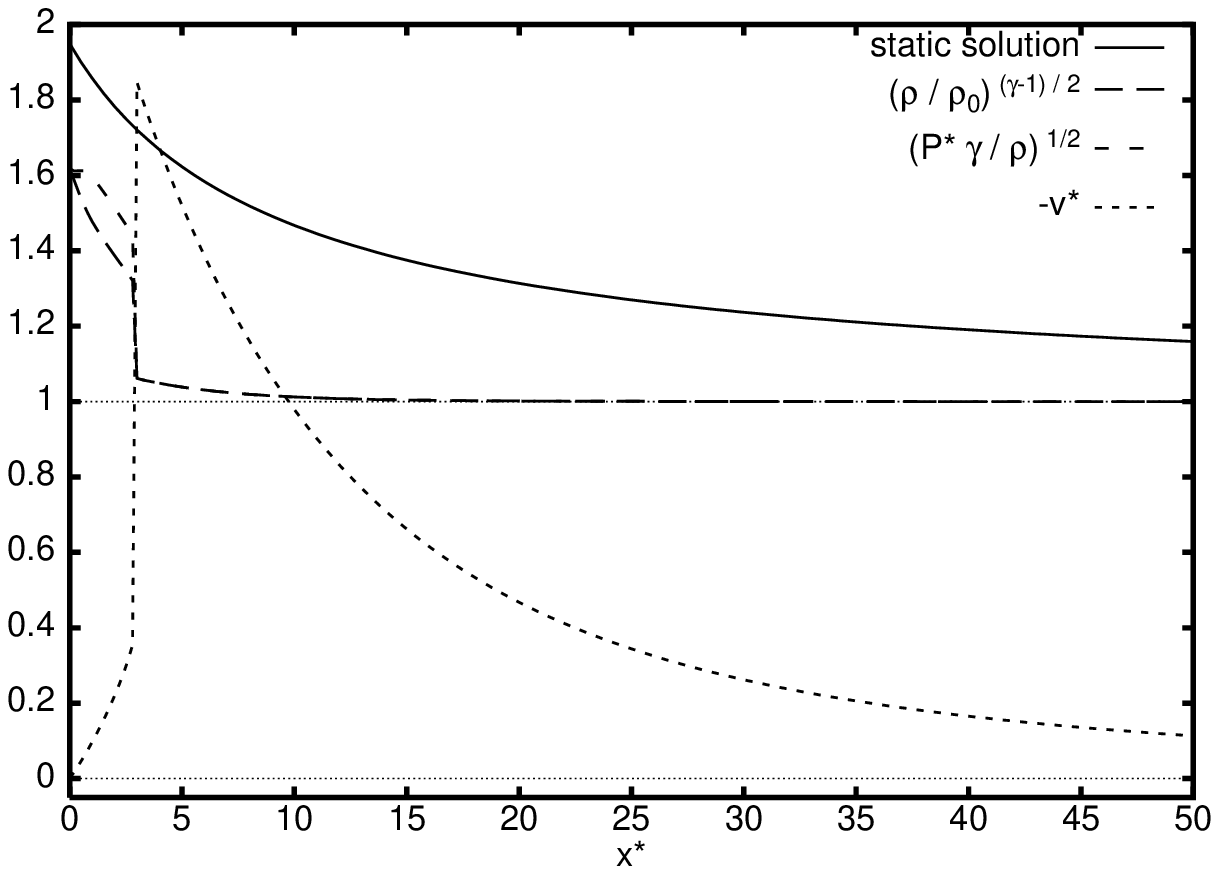,height=7.5cm,width=7.5cm}
\epsfig{figure=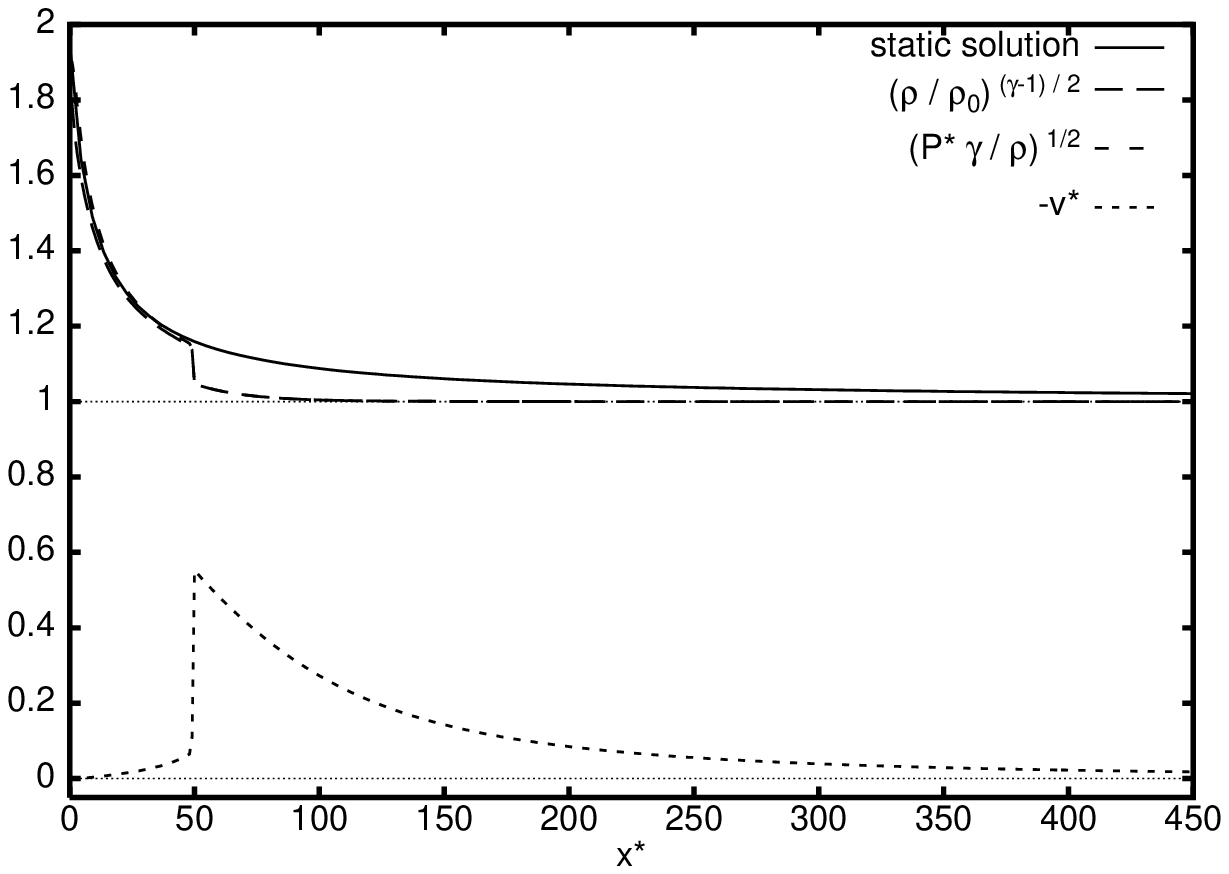,height=7.5cm,width=7.5cm}
\caption{\label{evol} Simulations for $R^{*}=7$: (Left) At $t^{*}=6$,
in a certain range of $x^{*}$, the speed of the downfalling gas $v^{*}$
is greater than the speed of sound in the media
$c^{*}=\sqrt{P^{*}\gamma/\rho}$ and a shock front is formed. At this
discontinuity the gas is no longer polytropic and $P^{*}\gamma/\rho$
is not given by $(\rho/\rho_0)^{\gamma-1}$. For comparison we include
the asymptotic solution or static solution given by equation
(\ref{law}).  (Right) At $t^{*}=70$ the speed is not supersonic, and
the gas is polytropic again. As the front moves away from the ball
surface, the density and pressure tend to the static solution.}
\end{center}
\end{figure}

The transition to the static state is best studied numerically.  Next
we show, for the case of varying gravity with $R^{*}=7$, two snapshots
of the spatial distribution of the main physical quantities
($v^{*}=v/c_0$, the speed of sound in the media $c^*=c/c_0=\sqrt{P^*
\gamma/\rho}$ with $P^*=P/c_0^2$, and $\rho/\rho_0$) as obtained from
the numerical integration explained in section \ref{num}, compared
with the values of the static solution given by equation (\ref{law}).
In Fig.\ \ref{evol} (left), we observe that the shock front appears as a
discontinuity of the main variables when the speed of the downfalling
gas is greater than the speed of sound (supersonic regime).  The shock
discontinuity decreases in magnitude once the gas speed $v^{*}$ is no
longer supersonic.  In Fig.\ \ref{evol} (right) we can see how the
numerical solution tends to the static solution. As the shock moves
away the gas behind slows down and finally becomes
still.

\section{Discussion}\label{discuss}

Summarizing, we have divided the space dimension (the radial distance $x$) 
into a near zone (near the ball's surface, where $x \ll R$) and a far zone
($x\gg R$).  In the near zone, we have obtained an exact solution for
short time and a self-similar regime for longer time and
radius. In the far zone, we have a self-similar regime for
relatively long time and radius. The parameter controlling the extent 
of both intermediate asymptotic regimes is $R^* = \R/R$. 
We have also obtained an exact static
state for very long times. Even though the regions of validity of the
analytic solutions occupy a small part of the plane $(x,t)$, the
numerical study provides an interpolation between them, 
in addition to verifying those solutions.  Therefore, the combined
solutions provide a good intuitive understanding of the entire
dynamics. The crucial feature is the formation and propagation of a
shock wave until its dissipation.

We have integrated numerically the equations of this dynamical system
with a robust scheme that, respecting the flux conservative structure
of the equations, is able to reproduce the dynamics in the
neighbourhood of the shock front. We observe that as soon as the gas
velocity is above the speed of sound in the media a sharp
discontinuity appears in all the relevant variables. In the case of
constant gravity the magnitude of this discontinuity increases up to a
constant value whereas the pressure and velocity increase
indefinitely. In the case of varying gravity, the highest
discontinuities are reached in the region $x_s \ll \R$, then the shock
front moves away from the ball and the magnitude of the discontinuity
decreases. It is interesting as well to remark that the shock wave is
particularly robust and holds for $x_s > \R$ (see the sharp
discontinuity in Fig.\ \ref{evol} (right)).

Since the propagation of the shock wave mostly takes place in the
regimes with negligible $P_0$ (if $R^* \gg 1$),
it is interesting to consider the
one-dimensional fluid dynamics equations for pressureless gas, namely,
the Burgers equation for the velocity (plus the continuity equation,
independent of it).  The Burgers equation includes viscosity $\nu$, but
in the limit of vanishing viscosity its only effect is to prevent the
velocity from becoming multivalued, giving rise to shocks
\cite{Frisch}.  Furthermore, the formation of shocks can be determined
analytically for {\em any} initial condition.  Therefore, it seems
that the intermediate asymptotic regimes should be given by the
corresponding exact solution of the Burgers equation, which is
trivial: the gas is in free fall everywhere but the velocity is
discontinuous at the ball's surface, in order that it be zero on
it. In other words, the shock is always at the ball's surface and does
not propagate. This corresponds to the pressureless {\em dust} being
deposited on the surface, where the density becomes infinite but
immediately jumps to its free-fall value.  Hence, we can appreciate
the importance of the dissipation associated to the shock wave:
whereas in the Burgers equation the kinetic energy of the free-falling
dust is simply ignored after it inelastically collides with the
ground, in the real intermediate asymptotic solutions it serves to
heat the gas below the shock wave, producing pressure where there was
none. This dissipation is connected with the breakdown of the 
thermodynamic equation (\ref{thermo}), which, in the case that 
this equation expresses entropy conservation, indicates 
that entropy is created at the shock.
The sharp rise of 
temperature and pressure behind the shock are clearly visible in the 
numerical simulations.

\subsection*{Acknowledgments}
We thank F.~Barbero, A.~Dom{\'\i}nguez, R.~G\'omez-Blanco, A.~Mancho and
J.~P\'erez-Mercader for conversations and comments. 
Furthermore, we thank the two referees of this paper for advise that has 
greatly contributed to improve it. 
The work of J.\ Gaite is supported by a ``Ram\'on y Cajal'' contract
and by grant BFM2002-01014, both of the Ministerio de Ciencia y
Tecnolog\'{\i}a. M.-P. Zorzano acknowledges an INTA fellowship for
training in astrobiology.

\appendix

\subsection*{Appendix: Similarity solution for the free fall in the field of
a point-like mass}

We describe here the ``Lagrangian solution'' of Eq.~(\ref{DE}) with $p=0$, 
that is, 
\be 
(3u-2)\,\xi u' + u^2 - u = -\xi^{-1}.  
\label{DE-p}
\ee 

Since the initial condition is given at $\xi \ra \infty$, it is 
convenient to make the change of variable ${\hat\xi} = \xi^{-1}$, 
transforming the equation into
\be
{du \over d{\hat\xi}} = {u^2 - u +{\hat\xi} \over (3u-2)\,{\hat\xi}}.
\label{eq}
\ee
The point (0,0) is a singular point of the ODE. Its 
solution is unspecified, unless we introduce an additional condition. 
The initial condition $v=0$ implies that 
$v=\sqrt{GM\over x}\,{\hat\xi}^{-1/2}u$
goes to zero as ${\hat\xi} \ra 0$. 
To impose it, it is best to linearize and solve
the ODE around (0,0): The general solution of 
\be
{du \over d{\hat\xi}} = {- u +{\hat\xi} \over -2{\hat\xi}}
\ee
is $u = -{\hat\xi} + C\,\sqrt{{\hat\xi}}$, so that the singular point 
is a node. 
Then, $\lim_{{\hat\xi}\ra 0}{\hat\xi}^{-1/2}u = C$. 
Hence, the particular solution of the nonlinear equation (\ref{eq})
in which we are interested is the only one that satisfies $u'(0) = -1$.

In the Lagrangian picture, the velocity is given by energy 
conservation as
\be
v = -\sqrt{2(E+ {GM\over x})}.
\label{v}
\ee
The initial condition that the particle is at rest implies that $E = -GM/x_0$.
Hence,
\be
u = {t\over x}v = -\sqrt{2\,{\xi}^{-1}(1- {x\over x_0})}\,, 
\label{u}
\ee
and to have a self-similar form we must express $x/x_0$ as a function of 
$\xi$.
In order to do it, we must solve the equation of motion (\ref{v}):
\be
t = -\int_{x_0}^x {ds\over \sqrt{2({GM/s}-{GM/x_0})}} =
-{x_0^{3/2}\over \sqrt{GM}} 
\int_1^{x/x_0} {d\s\over \sqrt{2({1/\s}-1)}}\,,
\label{int}
\ee
where we have introduced $\s = s/x_0$ to
write it in a self-similar form. 
It is clear now that $\xi^{-1/2} = F(x/x_0),$
which can be inverted to obtain $x/x_0$ as a function of $\xi$. 
The substitution of this function in Eq.\ (\ref{u}) yields
the solution of the ODE that satisfies the appropriate 
initial condition. 

One can calculate the integral in Eq.~(\ref{int}) by the change of variables 
$\s = \cos^2 \varphi$. This allows one to obtain the solution of the 
differential equation (\ref{DE-p}) in parametric form: 
Let 
\be
\cos^2 \alpha = x/x_0; \label{defa}
\ee
then,
\be
{\sqrt{GM}\, t\over {x_0}^{3/2}} =
{1\over \sqrt{2}}\,(\alpha + {\sin(2\alpha)\over 2}),  \label{ta}
\ee
and
\bea
\xi^{-1/2} &=& {1\over \sqrt{2}} \cos^{-3}(\alpha) \,(\alpha +
{\sin(2\alpha)\over 2}),  \label{xia}\\
u &=& -\cos^{-3}(\alpha) \,(\alpha + {\sin(2\alpha)\over 2})\,\sin\alpha,
\eea
where $\alpha \in (0,\pi/2)$. It is easy to verify that this parametric 
solution fulfills the initial condition $\lim_{\xi\ra\infty}u' = -1$.

In the Lagrangian formulation, the density is given by 
$$\r = \left(\partial x_0\over \partial
x\right)_{\! t} \frac{x_0^2}{x^2}\, \r_0,$$ where $x_0^2/x^2$ is a
spherical geometry factor. From Eqs.~(\ref{defa}), (\ref{ta}) and 
(\ref{xia}),
\be
r= {\r\over \r_0} = \frac{8\,\cos^{-3} (\a)}
  {9\,\cos (\a) - \cos (3\,\a) + 12\,\a\,\sin (\a)}, 
\ee
which, together with Eq.~(\ref{xia}), constitute the parametric equations
of the density.

\end{document}